\documentclass[10pt]{article}
\usepackage[utf8]{inputenc}
\usepackage[letterpaper]{geometry}
\usepackage{amsmath}
\usepackage[dvipsnames]{xcolor}
\usepackage{etoolbox}% EVIL
\usepackage{changepage}

\usepackage{marginfix} % necessary but possibly evil                                                 
\newcounter{marginnote}
\renewcommand{\footnote}[1]{\refstepcounter{marginnote}\textsuperscript{\themarginnote}\marginpar{\color{darkgray}\raggedright\footnotesize\textsuperscript{\themarginnote}#1\vspace{1ex}}} % see 1ex hack MAGIC

\renewcommand{\paragraph}[1]{\par\addvspace{1.5ex}\noindent\textbf{#1}~---}
\patchcmd{\thebibliography}% EVIL
  {\settowidth}
  {\setlength{\parsep}{1ex}\setlength{\itemsep}{0pt plus 0.1pt}\settowidth}
  {}{}
\newlength{\foo}
\newenvironment{wider}{\begin{adjustwidth}{0in}{-\foo}}{\end{adjustwidth}}
\sloppypar

\newcommand{\documentname}{\textsl{Note}}
\newcommand{\sectionname}{Section}
\newcommand{\foreign}[1]{\textsl{#1}}

\newcommand{\etc}{\foreign{etc.}}
\newcommand{\secref}[1]{\sectionname~\ref{#1}}
\newcommand{\noteref}[1]{Note~\ref{#1}}

\newcommand{\dd}{\mathrm{d}}
\newcommand{\e}{\mathrm{e}}
\newcommand{\bol}{\text{bol}}
\newcommand{\eff}{\text{eff}}
\DeclareMathOperator{\logten}{log_{10}}
\newcommand{\unit}[1]{\mathrm{#1}}
\newcommand{\m}{\unit{m}}
\newcommand{\pc}{\unit{pc}}
\newcommand{\kpc}{\unit{kpc}}
\newcommand{\au}{\unit{a.u.}}
\newcommand{\arcsec}{\unit{arcsec}}

\begin{document}\thispagestyle{empty}

\section*{\raggedright%
Magnitudes, distance moduli, bolometric corrections, and so much more}

\smallskip
\noindent\textbf{David W. Hogg}\footnote{% 
It is a pleasure to thank
Ivan Baldry (Liverpool),
Vasily Belokurov (Cambridge),
Rebecca Bernstein (GMT),
Michael Blanton (NYU),
Katie Breivik (Flatiron),
Aaron Dotter (Harvard),
Daniel Eisenstein (Harvard),
Will Farr (Stony Brook),
John Gizis (Delaware),
Gregory Green (MPIA),
Jim Gunn (Princeton),
Gerry Neugebauer (deceased),
Bev Oke (deceased),
Adrian Price-Whelan (Flatiron),
Hans-Walter Rix (MPIA),
and the Astronomical Data Group at the Flatiron Institute
for valuable discussions of these matters over the years,
and Sam Ward (Cambridge) for correcting an error.
I dedicate this to the memories of Arlo Landolt and Gerry Neugebauer, both of whom were cherished colleagues who devoted much of their careers to improving the precision and accuracy of fundamental astronomical measurements.}
{\footnotesize\par\noindent%
  \textsl{Center for Cosmology and Particle Physics, Department of Physics, New York University}\\
  \textsl{Max-Planck-Institut f\"ur Astronomie, Heidelberg}\\
  \textsl{Flatiron Institute, a division of the Simons Foundation}\par}

\bigskip
\paragraph{Abstract}
This pedagogical document about stellar photometry---aimed at those for whom astronomical arcana seem arcane---endeavours to explain the concepts of magnitudes, color indices, absolute magnitudes, distance moduli, extinctions, attenuations, color excesses, K~corrections, and bolometric corrections.
I include some discussion of observational technique, and some discussion of epistemology, but the primary focus here is on the theoretical or interpretive connections between the observational astronomical quantities and the physical properties of the observational targets.

\section{Introduction}\label{sec:intro}

Famously, an astronomical magnitude is negative-$2.5$ times the base-ten logarithm of a brightness or flux.
But that isn't quite right.
It is negative-$2.5$ times the base-ten logarithm of a \emph{ratio of two signals},
the numerator being the signal from a detector (and filter) measuring the electromagnetic radiation from the star of interest, and the denominator being the hypothetical (or actual) signal from that same detector (and filter) measuring a particular fundamental standard star (usually Vega, but sometimes another star like perhaps BD+17, or secondary standards, or hypothetical synthetic sources).
Magnitudes aren't just logarithmic, they are logarithmic measures of \emph{relative} brightness.

Astronomy is an \emph{observational} (as opposed to \emph{experimental}) science, and it is old.
This \documentname{} is aimed at physicists, who, reluctantly or otherwise, find that they must use astronomical observations in their research.
Or at theoretical astrophysicists, who can predict the physical properties of stars or galaxies or quasars, but must now predict the \emph{observed} properties of those same things, where those observed properties are curated by astronomers.
Even card-carrying observational astronomers\footnote{The author is a card-carrying observational astronomer.} can get confused by the relationships between magnitudes and luminosities, intensities, distances, and spectral energy distributions.
That's not embarrassing; the connections between in-practice observations and in-theory properties of stars (and galaxies so on) are complex.

Astronomers and non-astronomers alike love to make fun of the magnitude system as archaic and weird.
But it has an important epistemological role in astronomy, where controlled experiments are not possible,\footnote{You could imagine having a set of calibration sources orbiting the Earth; I will return to this idea at the end in \secref{sec:discussion}.} and the only calibration sources available for incompletely understood astronomical objects are \emph{other} incompletely understood astronomical objects:
Magnitudes are by their nature \emph{relative}, and relative in a purely empirical sense.
Of course there are questionable innovations, like the AB magnitude system \cite{ab}, which destroy the empirical relativity of the photometry system,\footnote{%
In what follows I will make some comments about the epistemological status of various photometric quantities.
The move made in the AB system away from a particular standard star and towards a counterfactual non-existent standard spectrum is helpful for theoretical interpretation, but bad for accuracy and legacy standardization.
In practice there isn't much difference, operationally, between AB and classic Vega-relative magnitudes, however, because in practice most projects are calibrating relative to secondary standards, as will be discussed in \secref{sec:practice}.
Furthermore, even in the Vega-relative system, the star Vega is often (in practice) replaced with a \emph{model for Vega} in calibration contexts, because Vega has some low-amplitude variability and a dust disk \cite{vegadust}.}
but the fundamental idea that a magnitude is based on a ratio of signals is sound and sensible:
By always involving ratios of empirical measurements, magnitudes permit extremely precise photometric measurements to be made in the face of large (historically enormous) absolute uncertainties about anything's true spectral density, and any detector's absolute sensitivity.

The magnitude system predates much of contemporary physics.
Indeed it predates quantitative natural science!\footnote{%
The numerical magnitude system predates natural science and yet the function $-2.5\,\logten()$ is a good fit to ancient by-eye numerical stellar classifications, as is the function $-\ln()$.
It's miraculous that intuitive human numbers come close to measurements in a natural-logarithm basis.
Presumably this reveals something about the human visual perception apparatus, and maybe also human language.}
Being ancient, the magnitude system, as astronomy has modernized into an intellectual enterprise that parallels and overlaps contemporary physics, entrains a great deal of jargon and esoterica.
These include distance moduli, K~corrections, color excesses, and bolometric corrections.
I will attempt to define and discuss all of these in in what follows.
This \documentname{} reinforces, repeats, and extends content to which I have contributed previously \cite{kcorrect}, and nothing here is new!
This is a purely pedagogical contribution.

Given that observational astrophysics is now technically sophisticated and extremely well equipped, why do we continue to use magnitude systems?
Obviously there are many answers to this question.
Some relate to the empirical precision considerations above.
But mostly it comes down to the point that astronomical photometric observations (especially of faint objects) involve integrations of stellar (or galaxy or quasar) radiation through broad (wavelength) bandpasses.\footnote{%
This argument owes a lot to discussions I've had over the years with Michael Blanton (NYU).}
There is no sense in which these photometric observations can be accurately summarized as spectral densities $f_\lambda$ at particular wavelengths $\lambda$; indeed our bandpasses are generally so wide that they don't resolve any of the interesting spectral features (atomic and molecular lines, for example).
The magnitude system is a compact and traditional way for astronomers to communicate precisely and concisely about these broad-band observations.

Because of this, there are conflicts in astronomy between the goal of measuring quantities that are simple and stable to calculate and theorize about, and the goal of measuring quantities that are simple and stable to observe and communicate about.
Theoretical simplicity and observational simplicity are desiderata in tension.\footnote{%
In addition to these concepts of simplicity, there are many contexts in which photometry needs to be precise in a relative sense but does not need to be precisely or accurately interpretable.
For example, extremely small exoplanet transits can be detected in photometry that is precise from epoch-to-epoch, even if the photometric system is not well defined in any other sense; for example \cite{kepler}.
For another, incredible detail in the color--magnitude diagram of stars is visible directly in apparent photometry \cite{hstcmd}.}
The magnitude system---at least the apparent magnitude system---is designed only with observational simplicity in mind.
Thus the goal here is to give the translations between these simple observational definitions and the theoretical quantities of interest.

Magnitudes and colors are based on ratios of detector signals, so they depend on the detector technology through which those signals are produced.
This means that any predictions of, or physical interpretations derived from, magnitudes and colors require detailed knowledge of the detector technology, and also the filter, optics, and atmosphere that lie between the detector and the astronomical sky.
There are subtleties involved, which I attempt to elucidate, below.
Suffice it to say that even something as apparently simple as the comparison of the brightnesses of two stars is not trivial, from a theoretical or fundamental-astronomy perspective.

I am not particularly interested in history here.
I am interested in the conceptual structure and relationships of quantities fundamental astronomy.
However, if you have interests in the history, there is at least one excellent book on the subject by Hearnshaw \cite{hearnshaw}.

\section{Apparent magnitude}\label{sec:mag}

The apparent magnitude $m_b$ of a star depends on a bandpass $b$ or its transmission function $R_b(\lambda)$, the star itself or its flux density $f_\lambda(\lambda)$, and a choice of fundamental standard star or standard flux density $f^{(0)}_\lambda(\lambda)$.
We'll say a lot more about the transmission function in \secref{sec:transmission};
for now let's imagine that the transmission function $R_b(\lambda)$ is a function that varies between 0 and 1.
It represents, at every wavelength $\lambda$ (or equivalently every frequency $\nu$) what fraction of the incident intensity at that wavelength (or frequency) incident \emph{on the top of the Earth's atmosphere} passes through the atmosphere and optical path and is detected or recorded by the detector.
We'll complexify that definition importantly in \secref{sec:transmission}.
In what follows, I will assume that the transmission function $R_b(\lambda)$ has been correctly estimated, incorporating atmosphere, optics, filter glass, and detector (all of which matter; more on this in \secref{sec:transmission}).

The flux density $f_\lambda(\lambda)$ has units of energy per time per area per wavelength.
It can be thought of as a luminosity density divided by an area, or it can be thought of as an intensity integrated over a solid angle.
It is an \emph{apparent} property of the star in the sense that it depends not just on the intrinsic properties of the target star, but also on the distance to the star, and on interstellar dust absorption and scattering.
Because the effect of the atmosphere is included in $R_b(\lambda)$ but the interstellar medium (say) is not, the relevant flux density in what follows is the flux density from the star \emph{falling on the top of Earth's atmosphere}.
The target star has a flux density $f_\lambda(\lambda)$ and the fundamental standard star or source (to which it will be compared to create the magnitude) has a flux density $f^{(0)}_\lambda(\lambda)$.

Armed with the transmission function $R_b(\lambda)$, the target-star flux density $f_\lambda(\lambda)$, and the fundamental standard-star flux density $f^{(0)}_\lambda(\lambda)$, the (asymptotically expected) magnitude $m_b$ as measured by a photon-counting device (such as a CCD) is given by the following
\begin{align}
    m_b &= -2.5\logten\frac{C_b}{C^{(0)}_b}\label{eq:mag}\\
    C_b &= \int \frac{f_\lambda(\lambda)}{(h\,c/\lambda)}\,R_b(\lambda)\,\dd\lambda\label{eq:counts}\\
    C^{(0)}_b &= \int \frac{f^{(0)}_\lambda(\lambda)}{(h\,c/\lambda)}\,R_b(\lambda)\,\dd\lambda ~,
\end{align}
where $C_b$ and $C^{(0)}_b$ are numbers proportional to the expected \emph{photon counts} detected per unit time in an exposure of the target star and of the standard star,
and the all-important factor of $h\,c/\lambda$ turns energy flux density into photon flux density.
Because the magnitude is a logarithm of a ratio of (dimensionally identical) signals, the magnitude is a dimensionless quantity, by construction.

If you prefer to work in frequency units, you can use the flux density $f_\nu(\nu)$ that has units of energy per time per area per frequency, and the expected counts equations become
\begin{align}
    C_b &= \int \frac{f_\nu(\nu)}{h\,\nu}\,R_b(\nu)\,\dd\nu \label{eq:countsnu}\\
    C^{(0)}_b &= \int \frac{f^{(0)}_\nu(\nu)}{h\,\nu}\,R_b(\nu)\,\dd\nu ~,
\end{align}
where now the factor of $h\,\nu$ turns energy flux density into photon flux density, but everything else is the same.\footnote{%
  Sometimes the integral \eqref{eq:countsnu} is written as $$\int f_\nu(\nu)\,R_b(\nu)\,\dd\ln\nu$$ but I don't like this because writing $\dd\nu/\nu$ as $\dd\ln\nu$ obscures the point that the factor of $1/\nu$ comes from the energy-per-photon factor $h\,\nu$.}
The two different kinds of flux densities $f_\lambda(\lambda)$ and $f_\nu(\nu)$ and the two transmission functions $R_b(\lambda)$ and $R_b(\nu)$ are related by
\begin{align}
    \nu\,f_\nu(\nu) &= \lambda\,f_\lambda(\lambda) \mbox{~~s.t.~} \lambda = \frac{c}{\nu}\\
    R_b(\nu) &= R_b(\lambda) \mbox{~~s.t.~} \lambda = \frac{c}{\nu} ~.
\end{align}

The important consequences of the magnitude definition \eqref{eq:mag} are as follows:
If your target star is ``fainter'' than the fundamental standard source in the bandpass $R_b(\lambda)$---meaning that if it delivers a smaller signal in your device in this bandpass---then it will have a positive magnitude.
The magnitude gets larger as the target star gets fainter.
If the magnitude is zero (this situation is rare!), the target star is identical in apparent brightness to the fundamental standard star, given your device and bandpass.

Although the definition \eqref{eq:mag} is in terms of the fundamental standard star with flux density $f^{(0)}_\lambda(\lambda)$, in many cases the actual mechanism of calibration is comparison not with the fundamental standard, but with some secondary standard that has been calibrated to the fundamental.
If we have a secondary standard star with flux density $f^{(1)}_\lambda(\lambda)$ and (known) apparent magnitude $m_b^{(1)}$ in bandpass $R_b(\lambda)$, the magnitude $m_b$ of the target star can be expressed as a different ratio of signals:
\begin{align}
    m_b - m_b^{(1)} &= -2.5\logten\frac{C_b}{C^{(1)}_b}\label{eq:secondary}\\
    C^{(1)}_b &= \int \frac{f^{(1)}_\lambda(\lambda)}{(h\,c/\lambda)}\,R_b(\lambda)\,\dd\lambda ~,
\end{align}
where we have replaced the fundamental standard $f^{(0)}_\lambda(\lambda)$ with the secondary standard $f^{(1)}_\lambda(\lambda)$, and adjusted the magnitude to be a magnitude difference.
We'll return to this briefly in \secref{sec:practice}.

In classical astronomical parlance, the magnitude $m_b$ can be called an \emph{apparent magnitude}, where the modifier ``apparent'' indicates that it is a purely observational quantity, observed at the Earth.
The apparent magnitude has the simple property that it can be measured without additional assumptions, but it has the non-simple property that its theoretical prediction or understanding involves not just the physical (intrinsic) properties of the star; it also involves the distance to the star, interstellar extinction, and other effects, all of which will be discussed below.
From an epistemological standpoint, the apparent magnitude is extremely fundamental, since it can be measured by taking ratios of signals measured in detector--telescope setups without many additional assumptions.

At the risk of being extremely repetitive:
The point of this definition \eqref{eq:mag} of the apparent magnitude is that it is very simple to very precisely \emph{measure} the magnitude.
The definition is given in terms of integrals of the spectrum $f_\lambda(\lambda)$ of the source, and the spectrum $f^{(0)}_\lambda(\lambda)$ of the fundamental standard star, and the transmission function $R_b(\lambda)$ of the bandpass.
But---and this couldn't be more important---you don't need to \emph{know} any of these functions in order to measure the magnitude.
Magnitudes existed when the only instruments observing the sky were human eyes, and knowledge of how the stars shine was considered a theoretical impossibility (and maybe also an affront to God).
We'll say more about in-practice measurement of magnitudes in \secref{sec:practice}.
But in most of this \documentname{} our concern is with the relationships between the observed magnitudes (or their expectations\footnote{%
Strictly speaking the expressions in this \sectionname{} are expressions for the \emph{expectation} of the apparent magnitude. In detail, any measured magnitude involves taking the ratio of noisily measured quantities. I'll say a little more about this in \secref{sec:practice}.})
and the theoretical quantities, such as flux densities and lumninosity densities, not because these theoretical quantities are necessary to \emph{measure} magnitudes, but instead because they are necessary to \emph{understand} magnitudes.

What's written here (and in most of what follows) is very biased towards the Vega-relative magnitude system.
In this system, Vega ($\alpha$~Lyrae) is the fundamental standard star; it has zero magnitude in all bandpasses.
The next-most common system is the AB magnitude system \cite{ab}, which makes use not of a fundamental standard star as the reference source, but of a hypothetical non-existent source with a spectrum that is flat in $f_\nu(\nu)$.
That makes AB magnitudes not directly observable but only observable through absolutely-calibrated secondary standards \cite{ab}.
The AB magnitudes are thus easier to interpret theoretically, but harder to measure accurately, and more subject to revision, as our understandings of bandpasses, transparencies, efficiencies, and the spectrophotometry of secondary standards evolve.
Full discussion of these issues are outside our scope here, but suffice it to say that the theoretical simplicity of the AB magnitude system comes at a significant observational cost.

\section{Transmission function}\label{sec:transmission}

The most important thing defining the magnitudes in band $b$ is the total combined transmission function $R_b(\lambda)$.
That transmission function has contributions from multiple causes or elements in the optical path from the target star to the telescope.
In most cases, the most important contribution comes from the colored glass filter deliberately placed in the optical path of the light, but there are also important contributions to the transmission function from the atmosphere (which absorbs and scatters light of different wavelengths very differently) and from the detector hardware itself (which is usually at least slightly differently sensitive to photons of different wavelengths, and has some long-wavelength cutoff).
There are also small (we hope) contributions from the telescope optical surfaces,
and there are additional complexities arising from the fact that the atmosphere changes with time, and different observations are made through different atmospheric depths, but I will do no more than comment on these things below in \secref{sec:practice}.
Symbolically, the transmission function can be written as a product of terms:
\begin{align}
    R_b(\lambda) &= R_\text{atmosphere}(\lambda)\,R_\text{optics}(\lambda)\,R_\text{filter}(\lambda)\,R_\text{detector}(\lambda)~,
\end{align}
where the atmosphere term depends on airmass and conditions, and the filter term is the only part that the observer directly controls.

The way that the transmission function $R_b(\lambda)$ is used above in \eqref{eq:counts} demonstrates, dimensionally,
that the value of the function $R_b(\lambda)$ is the expected contribution of a \emph{photon} of wavelength
$\lambda$ incident on the top of the atmosphere to the signal $C_b$ read out from the detector (in some sense, it has units of signal per photon).
If the detector is a photon-counting device, then the function $R_b(\lambda)$ varies between zero and unity, and it can be measured (in principle) by measuring relative intensities at different points in the optical path from outer space to the detector.
However, if the detector is a bolometer or an energy-integrating device, the function $R_b(\lambda)$ has to be inferred
with a bit more care; it involves an extra factor of $h\,c/\lambda$ to deliver, again, the expected contribution of
a \emph{photon} of wavelength $\lambda$ to the signal read out by the detector.

That is, if you have old-school technology, or very new-school technology, or you work in the far infrared, you might conceivably have some kind of \emph{bolometer}, or energy-integrating detector.
In this case, you get a different expression for the apparent magnitude $m_b$, as follows
\begin{align}
    m_b &= -2.5\logten\frac{S_b}{S^{(0)}_b}\\
    S_b &= \int f_\lambda(\lambda)\,S_b(\lambda)\,\dd\lambda\\
    S^{(0)}_b &= \int f^{(0)}_\lambda(\lambda)\,S_b(\lambda)\,\dd\lambda~,
\end{align}
where now $S_b$ and $S^{(0)}_b$ are numbers proportional to the bolometer signal detected in an exposure of some duration from the target star and the fundamental standard star.
These expressions for $S_b$ and $S^{(0)}_b$ are integrated energies (or really energies per area per time), so they do not involve any factors of $h\,c/\lambda$.
But very importantly to our message here, if $R_b(\lambda)=S_b(\lambda)$, the photon-counting and bolometer magnitudes \emph{are not the same}!
These two magnitudes the same \emph{if and only if}
\begin{align}
    R_b(\lambda) &= \frac{h\,c}{\lambda}\,S_b(\lambda)~.\label{eq:StoR}
\end{align}
From here on, in this document, It will be assumed that the detector is of the photon-counting variety, not the energy-integrating variety.
But everything is completely technology-independent if \eqref{eq:StoR} holds.
What isn't technology-independent is that if you have a piece of colored glass that is, say, your $V$-band filter glass, that you use on your photon-counting CCD imager, and you take it to your bolometer-equipped photometer, and you use that exact same piece of colored glass, you get a different $V$-band apparent magnitude, solely because the photon-counting and energy-integrating devices do different integrals of the incident intensity.

One point worthy of note here, perhaps, is that if you measure the transmission of a piece of colored filter glass in the usual way---by comparing the intensity transmitted to the intensity incident---you obtain the multiplicative contribution of that filter glass to $R_b(\lambda)$, if you are going to put it in front of a photon detector.
That is, since most imagers and photometers are photon-counting devices, in most cases when you are measuring a filter curve in a piece of glass, you are measuring a multiplicative contribution to $R_b(\lambda)$.
However, if you are going to put that filter glass in front of a bolometer, what you measure is the multiplicative contribution of that filter glass to $S_b(\lambda)$.

There are multiple terms used for the transmission function $R_b(\lambda)$ in the literature.
The NASA \textsl{Hubble Space Telescope} documentation calls the transmission function $R_b(\lambda)$ the ``integrated system throughput'' \cite{acs}.
The ESA \textsl{Gaia} documentation calls it the ``transmission'' or ``response'' function \cite{jordi}.
The \textsl{Sloan Digital Sky Survey} documentation calls it the ``response'' function \cite{doi}.
Sometimes these projects also compute ``effective wavelengths'' for the bandpasses, but I don't think this is very useful, so I am putting it out of scope.\footnote{The concept of an ``effective wavelength'' is not useful for two related reasons. The first is (as noted in \secref{sec:intro}) most photometric bandpasses are so broad that it is not responsible to summarize the photometry in terms of flux densities at particular wavelengths; they are integrals over wavelength. The second is that any definition of effective wavelength essentially depends on the underlying spectrum of the target star, so it isn't really a property of the bandpass itself.}

One amusing aspect of the transmission function is that it involves a term that represents the sensitivity of the detector, and multi-pixel imaging detectors (like CCDs) are made up of millions of (at least slightly) heterogeneous pixels.
That means that, technically, even when a well-defined filter is in place, every pixel of a real-world imaging system has a slightly different transmission function, and that the effective bandpass through which an apparent magnitude is measured is a (maybe very weak) function of position in the detector.
That's interesting, but out of scope here.\footnote{%
It is amusing to note here that one of the high-cost items in the Rubin Observatory budget has been the filters, which are designed to be extremely consistent across the (enormous) focal plane \cite{lsstfilters}.
The choice to spend here is (retrospectively) debatable, since the detector pixels themselves (plus variations in atmosphere, airmass, and so on) will lead to bandpass variations over the focal plane (and time), which will have to be accounted for in any precise analysis no matter how good the filter glass is.}

In addition to every pixel having a slightly different transmission function, every exposure taken (from ground-based hardware, anyway) will have a slightly different transmission function too.
This is because the atmosphere is not a fixed object, but changes with pressure, temperature, (very importantly) humidity, dust content, cloud cover, and airmass.\footnote{Airmass is defined to be the length of the optical path through the atmosphere, normalized to the optical length at the zenith. It is unity when the telescope is pointed directly at the zenith, and it depends (approximately) on the zenith angle as the secant.\label{note:airmass}}
Even if the atmosphere is unchanging at the relevant wavelengths, the airmass dependence of the transmission function makes it true that every observation in a sequence will have a different transmission function (as the target star rises and sets).
In principle, extremely precise photometric projects will have to track and account for these atmospheric changes.
They matter most at the blue edge of the atmospheric transmission, and in infrared windows that are bounded by atmospheric water features.
The airmass dependence will reappear below in \secref{sec:practice}.

\section{Color index}\label{sec:color}

Much of astronomy is the study of star (or galaxy or quasar or supernova or \etc) spectra or spectral-energy distributions.\footnote{%
It is with apologies on behalf of my communities that I comment here that astronomers use the term ``spectral-energy distribution'' to mean the spectrophotometrically calibrated flux density from a source as a function of wavelength or frequency, usually at low spectral resolution.}
If you only have imaging devices at your disposal---which make photometric measurements---and not a spectrograph, then your spectral information comes from comparing the brightness of the object in different bandpasses, each of which has different wavelength coverage.
And, if you accept the relative properties of the magnitude system (that is, that all measurements are made relative to a standard star), then the spectral information delivered by multi-band photometry is, fundamentally, information about \emph{differences in the amplitude and shape} between the spectrum of your target star and the spectrum of the standard star.

The classical optical bandpasses are maybe $U,B,V,R,I,J,H,K$ spanning from the ultraviolet cutoff of the atmosphere to the longest wavelengths practicable with array detectors in windows of atmospheric transparency in the infrared.\footnote{%
Sometimes people object to my saying that $J, H, K$, which are in the near infrared, are ``optical'' bandpasses. But my position (inherited from Gerry Neugebauer, a founder of infrared astronomy), is that ``optical'' is the name of a set of techniques, while ``ultraviolet'', ``visible'', and ``infrared'' are electromagnetic wavelength ranges. So the \textsl{Spitzer Space Telescope} \cite{spitzer}, for example, was an optical instrument observing in the infrared.}
There are lots of other bandpasses in use (including say $u,g,r,i,z,y$), and from space the wavelength range is enormously expanded from the far-ultraviolet through the mid-infrared.\footnote{%
You can detect a slight prejudice against gamma-ray, X-ray, submillimeter, and radio here. High-energy and radio astronomy never really used these definitions.}
For each of these bandpasses, if you want to interpret theoretically the spectral information it provides, it is critical to have some estimate of the transmission function $R_b(\lambda)$, as I discussed in \secref{sec:mag}.

A color index---or more usually just a ``color'' these days---is a difference of two magnitudes, measured through two different bandpasses.
Why a difference of two magnitudes?
A difference of magnitudes is a difference of logarithmic signal ratios, which is like a logarithm of a ratio of signal ratios.
Imagine, for example, that the color in question is the difference $B-V$, or the difference between the apparent magnitude $m_B$ of the target star measured in the $B$ band and the apparent magnitude $m_V$ of the target star measured in the $V$ band.
This color---or magnitude difference---can be rearranged as follows:
\begin{align}
    B-V &= -2.5\logten\frac{C_B}{C^{(0)}_B} + 2.5\logten\frac{C_V}{C^{(0)}_V}\\
        &= -2.5\logten\frac{C_B/C_V}{C^{(0)}_B/C^{(0)}_V}\label{eq:color} ~,
\end{align}
where $C_B$ is the expected detector signal for the target star through the $B$ bandpass (defined by \eqref{eq:counts} but using the $B$-band transmission function $R_B(\lambda)$),
$C^{(0)}_B$ is the expected detector signal for the fundamental standard star through the $B$ bandpass,
$C_V$ is the expected detector signal for the target star through the $V$ bandpass (defined by \eqref{eq:counts} but using the $V$-band transmission function $R_V(\lambda)$), and
$C^{(0)}_V$ is the expected detector signal for the fundamental standard star through the $V$ bandpass.
Expression \eqref{eq:color} shows that the color index is related to the ratio of the two signals through the two bandpasses, relative to that same ratio for the standard star.
Just like with the apparent magnitudes, we also often measure colors using secondary standards.
In this case the color index is written as a difference of differences
\begin{align}
    (B-V) - (B-V)^{(1)} &= -2.5\logten\frac{C_B}{C^{(1)}_B} + 2.5\logten\frac{C_V}{C^{(1)}_V} ~,
\end{align}
where $C^{(1)}_B$ and $C^{(1)}_V$ are the signals from the secondary standard, which has a known color $(B-V)^{(1)}$ in the system defined by the fundamental standard.

It is important in astronomical tradition that color indices are always defined with the shorter-wavelength bandpass first (positive) and the longer-wavelength bandpass second (negative).
So, for example, it is always $(B-V)$ and never $(V-B)$.
It is worth noting, however, that there can be ambiguous cases, such as when the two bandpasses overlap and have different widths.
But whenever it is unambiguous, in order to avoid confusion, the color index should always be defined as the shorter-wavelength bandpass magnitude minus the longer-wavelength bandpass magnitude.

The important consequences of these definitions (and the bandpass order convention) are as follows:
If your target star is ``redder'' than the fundamental standard star in the pair of bandpasses $R_1(\lambda)$ and $R_2(\lambda)$---meaning that if it delivers a larger relative signal (relative to the standard star) in the longer-wavelength bandpass than it does in shorter-wavelength bandpass, then it will have a positive color index.
The color gets larger as the target star gets redder (in this very specific sense of ``redder'').
If the color index is zero, the target star is consistent with being identical in spectral shape to the fundamental standard star, given your device and the chosen pair of bandpasses.
If the color index is negative, the target star is bluer than the standard.

\section{Distance modulus and absolute magnitude}\label{sec:absmag}

I have emphasized that magnitudes are \emph{apparent} properties of the star.
That means that---if you want to predict them or interpret them---your expectations don't only depend on the intrinsic, local, physical properties of the target star;
they also depend on the distance to the star (and, below, we will find that they depend on interstellar attenuation, and even redshift, when things are moving fast).
This \sectionname{} will begin with the simplifying assumption that the interstellar attenuation and the redshift of the target star are both negligible.

Historically, because it was so important to the astronomical community that magnitudes be \emph{relative} to a standard star, the traditional magnitude-related definition of luminosity is also constructed to be technically relative:
The \emph{absolute magnitude} of a star is defined to be the counterfactual magnitude it \emph{would have had} if the star had been situated $10\,\pc$ from the Solar System (or really the observer).\footnote{%
The parsec ($1\,\pc$) is a parallax arcsecond or about $206265\,\au$
The absolute-magnitude distance of $10\,\pc$ is about $3.086\times 10^{17}\,\m$.
It corresponds to an Earth-orbit astrometric parallax of of $0.1\,\arcsec$.}
History is not at stake in this \documentname, but historically this definition comes from the fact that the bright star Vega ($\alpha$~Lyrae) is about $10\,\pc$ away, and Vega was the traditional standard star of the consensus photometric bandpasses.\footnote{%
Vega is now thought to be at a distance of about $8\,\pc$, but the magnitude system is established and the $10\,\pc$ definition is considered immutable.
But of course this shows the arbitrariness of trying to make an intrinsic luminosity system that is relative:
As your beliefs about the distance to the fundamental standard star change, do you update the definition of the absolute magnitude, or do you live with the fact that the fundamental standard star no longer has an absolute magnitude of zero?
Our community has chosen the latter.
In a Vega-relative photometric system, Vega has an apparent magnitude of zero (by definition), but an absolute magnitude slightly greater than zero (because it is closer to us than $10\,\pc$).}
The absolute magnitude is a measure of the intrinsic luminosity of the target star (as opposed to the apparent brightness) because if all the stars were at the same distance from us (and unaffected by interstellar attenuation), their relative brightnesses would be in proportion to their intrinsic luminosities.

An absolute magnitude $M_b$ is related to an apparent magnitude $m_b$ through a \emph{distance modulus} $DM$, which is a logarithmic measure of the distance to the star:\footnote{Note that the distance modulus is traditionally denoted by the two-letter symbol ``$DM$'' and is not in any sense a product of a $D$ times an $M$.}
\begin{align}
    m_b &= M_b + DM\label{eq:absmag}\\
    DM &= 5\,\logten\frac{D}{(10\,\pc)}\label{eq:DM}\\
       &= -5\,\logten\frac{\varpi}{(0.1\,\arcsec)} ~,
\end{align}
where $D$ is the distance to the star,
$\varpi$ is the astrometric parallax,
and I have assumed (for now) that there is no interstellar attenuation (but see \secref{sec:extinction}).
The distance-modulus expression \eqref{eq:DM} invoves a $+5$ and not a $-2.5$ because the apparent brightness is inversely proportional to the square of the distance to the star.

The consequences of these definitions are as follows:
The distance modulus is measured in the same units as magnitudes, and it increases with distance.
$DM=0$ at $D=10\,\pc$, and $DM=10$ at $D=1\,\kpc$ (for example).
The absolute magnitude $M_b$ of the target star through a transmission function $R_b(\lambda)$ is near zero if its intrinsic luminosity is similar to the intrinsic luminosity of the fundamental standard star, as measured by photon luminosity in the bandpass.
The absolute magnitude is positive if the target star---placed counterfactually at a distance of $10\,\pc$---would be fainter than the fundamental standard star, and negative if the target star would be brighter.

One absurdity of cosmology is that absolute magnitudes are often used for cosmologically distant sources like quasars and galaxies, which could never be observed from a distance of $10\,\pc$---they are far, far larger in size than $10\,\pc$!
But the numerical definition of the distance modulus works for even enormous distances, so absolute magnitudes can be defined for extremely large, luminous sources, even if they are absurd when thought of as being counterfactually observed at a distance of $10\,\pc$.
Galaxies and quasars obtain absolute magnitudes that are negative (of course!) with large absolute values.
When thinking about a cosmologically distant source, the recession velocity will typically be large (even relativistic), and the concept of distance gets complexified \cite{distances};\footnote{%
Technically cosmology is not required to complexify the concept of distance. Even in special relativity, when a source is moving relativistically, there will be a difference between, say, the angular-diameter and luminosity distance to that source. This phenomenon isn't particularly cosmological.}
the distance $D$ in \eqref{eq:DM} is technically the \emph{luminosity distance}; I will say more about this in \secref{sec:kcorrect}.

Because absolute magnitudes depend on knowledge of distance---they can't be measured without a distance estimate---they are less purely observational than apparent magnitudes.
That puts them in a weaker epistemological class than apparent magnitudes and color indices.
That said, in the age of the ESA \textsl{Gaia} Mission \cite{gaia}, distances to stars are being measured in large numbers, and with minimal assumptions.
That is, it is possible to have a pretty pure measurement of stellar distance, not strongly dependent on complex physical assumptions.
However, the epistemology of magnitude-related quantities will get weaker as we progress, below.

\section{Bolometric correction}\label{sec:bc}

In many astronomical projects, the purpose of performing precise photometry is to test or use theoretical models of stars.
Stellar models can, in many cases, predict stellar luminosities.
These luminosities are often total, or \emph{bolometric}, luminosities, in the sense that they represent the total energy output of the star, integrating over all wavelengths.
Strictly speaking, stars don't have precisely well-defined luminosities, and they don't have precisely defined radii (because of complexities of the photosphere).
However, if a star \emph{did} have a well defined bolometric luminosity $L$ and photospheric radius $r$, then the star would have an effective temperature $T_\eff$ defined by
\begin{align}
    L &= 4\pi\,r^2\,\sigma\,T_\eff^4 ~,
\end{align}
where $\sigma$ is the Stefan--Boltzmann constant.
The bolometric luminosity can be computed if you know the distance $D$ to the star\footnote{%
Oddly the distance $D$ here is not precisely the luminosity distance, but the relevant distinction is pedantic when looking at nearby stars moving at non-relativistic speeds.
The reason is that the luminosity distance converts between $\lambda\,f_\lambda$ and $\lambda\,L_\lambda$; if converting between $f_\lambda$ and $L_\lambda$ there is a factor of $[1+z]$ to account for the ``per wavelength'' properties of those spectral densities.
This is pedantry, however, because the bolometric corrections in question in this \sectionname{} are never (or rarely) used in cosmological or relativistic contexts.}
and the full spectral flux density $f_\lambda(\lambda)$:
\begin{align}
    L &= \int L_\lambda(\lambda)\,\dd\lambda\label{eq:L}\\
    L_\lambda(\lambda) &= 4\pi\,D^2\,f_\lambda(\lambda)\label{eq:L_lambda}~,
\end{align}
where $L_\lambda(\lambda)$ is the spectral luminosity density (the luminosity equivalent of the spectral flux density).
In general you can only know this function $L_\lambda(\lambda)$ with a theoretical model of a stellar photosphere, coupled to a model of a stellar interior; it is not a direct observable.
This means that bolometric absolute magnitudes and bolometric corrections are epistemologically weaker than even absolute magnitudes, which are themselves epistemologically weaker than apparent magnitudes.
Absolute bolometric magnitudes and bolometric corrections require physical models of stars for their estimation.

The absolute bolometric magnitude $M_\bol$ of a star of bolometric luminosity $L$ is defined by
\begin{align}
    M_\bol &= -2.5\,\logten\frac{L}{L^{(0)}}\\
    L^{(0)} &= 4\pi\,D_{(0)}^2\int f^{(0)}_\lambda(\lambda)\,\dd\lambda \\
            &= \int L^{(0)}_\lambda(\lambda)\,\dd\lambda ~,
\end{align}
where $L^{(0)}$ is the bolometric luminosity of the fundamental standard star,
and $D_{(0)}$ is the distance to the fundamental standard star.
For comparison of data to theoretical models of stars, it is useful often to compare the data and models in terms of bolometric luminosities, or absolute bolometric magnitudes.
Bolometric magnitudes are more easily interpretable, theoretically, than their absolute magnitudes $M_b$ in any particular bandpass $R_b(\lambda)$.
Thus it was natural for astronomers to define a bolometric correction $BC_b$ which converts between the two:
\begin{align}
    M_b &= M_\bol - BC_b\label{eq:BC}\\
    m_b &= M_\bol - BC_b + DM ~,
\end{align}
where again I have assumed that interstellar attenuation is negligible.
The sign of the bolometric correction was defined this way, which (in my humble opinion) is wrong,\footnote{All other corrections have the sign convention that the apparent magnitude is found by adding in positively signed corrections to the absolute magnitude.} but it is set now.
Because the bolometric correction $BC_b$ involves a ratio of (integrals of) luminosities of stars, and not a ratio of apparent brightnesses, the bolometric correction is a distance-independent (and, we will see below in \secref{sec:extinction}, interstellar-absorption-and-scattering-independent) property of the target star.

I commented above that the bolometric magnitudes are epistemologically weak, because they involve stellar models for their computation.
Sometimes astronomers think of bolometric magnitudes as being somehow direct observables in some cases.
After all, the target star can be observed through a set of bandpasses that span the full spectral range in which the target star emits significantly.
However, the bolometric magnitude of the target star is not a direct observable, even in this ideal case.
After all, the photometric information coming from each individual band involves the ratio of the signal from the target star and from the fundamental standard star (or a secondary standard).
Thus it isn't possible to reconstruct, even from a large set of photometric measurements in a set of bandpasses that span the full spectral range, the bolometric magnitude of the star without an accurate spectral model for the standard star.

One of the extremely unpleasant things about bolometric corrections is that their definition \eqref{eq:BC} mixes up luminosity integrals with energy-per-time units and telescope-signal integrals with photon rate units.
That creates even more opportunities for investigators to get confused about what's being measured and how.
This problem can't be fixed with changes to conventions though:
It really is photon rates that are measured in (contemporary) photometry, and it really is total luminosities (energies per time) that are computed by theoretical models of stars.

The consequences of these definitions are as follows:
If a star has a bolometric luminosity greater than the fundamental standard, it will have a negative absolute bolometric magnitude; if it has a smaller bolometric luminosity it will have a positive absolute bolometric magnitude.
The bolometric correction is a bit harder to reason about.
The bolometric correction can be positive or negative.
If the stars are close to being black-bodies, the rough expectations are as follows:
The target star will have a vanishing bolometric correction if the star has the same spectral shape as the fundamental standard star, or if the bandpass $R_b(\lambda)$ is beautifully and coincidentally centered on some (slightly hard to write down) natural wavelength related to the peaks in the spectral density coming from the target star and the fundamental standard star.
The target star will have a positive bolometric correction if it is redder (cooler) than the fundamental standard star and it is being observed through a bandpass with support at sufficiently long wavelengths.
Alternatively, the target star can also have a positive bolometric correction if it is bluer (hotter) than the fundamental standard star and it is being observed through a bandpass with support at sufficiently short wavelengths.
There are many other situations possible depending on the shapes of the stellar spectra and the specific locations and widths of the bandpasses.
There are common misconceptions about the sign of the bolometric correction, but that's not surprising, since these conditions are non-trivial.

The Sun is the most well-understood star in the Universe (by humans, anyway), so it is common to reference bolometric quantities to the Sun instead, not to the fundamental standard star Vega.
In this case the absolute bolometric magnitude itself becomes a sum of two terms
\begin{align}
    M_\bol &= M_{\bol,\odot} - 2.5\,\logten \frac{L}{L_\odot}~,
\end{align}
where $M_{\bol,\odot}$ is the absolute bolometric magnitude of the Sun, and the second term is like a bolometric absolute magnitude but referenced to the Sun instead of the fundamental standard.\footnote{%
Although the Sun and Vega are about as well understood as stars can be, theorists are not always on the same page with regard to the Vega-relative Solar bolometric magnitude. It is important enough to astrophysical practice that the IAU adopted a reference value with resolution 2015.B2 to make the literature more definite \cite{2015b2}. Because of this resolution, most bolometric magnitudes are referenced to the Sun now, not Vega.}
This might seem strange, but the bolometric correction of the fundamental standard is not zero, even in the Vega-relative system, and thus it is not better than the Sun, conceptually, and the Sun is better understood, theoretically.

Since the absolute magnitude $M_b$ in bandpass $R_b(\lambda)$ is related to the apparent magnitude $m_b$ by a distance modulus offset, it is possible to define in parallel a bolometric apparent magnitude $m_\bol$ in parallel to the bolometric absolute magnitude $M_\bol$ as
\begin{align}
    m_b &= m_\bol - BC_b ~,
\end{align}
but perhaps these aren't used as much as the bolometric absolute magnitudes?

\section{Interstellar extinction and color excess}\label{sec:extinction}

At various points above it has been assumed that there is no significant interstellar attenuation.
But of course for the majority of stars in our Galaxy, this is not true!
Intervening interstellar dust (and gas, but mainly dust\footnote{Astronomers call this interstellar material ``dust'' but it is really a range of different kinds of material including graphite, silicates, hydrocarbons, and other complex molecules, plus particles of solid-state materials in a wide range of sizes. It might be better called ``sand and smoke''? It is absolutely not at all what general relativists call ``dust''! And it is also not what you call dust when you sweep your floors (is that really mainly dead skin cells?).}) absorbs and scatters light from the stars, and preferentially at shorter wavelengths.
For this reason, dust not only attenuates stellar brightnesses but also shifts stellar colors to the red.

Above in \eqref{eq:L_lambda} the relationship between the spectral flux density $f_\lambda(\lambda)$ and the spectral luminosity density $L_\lambda(\lambda)$ was defined in the context that there is no dust attenuation.
If there is intervening interstellar dust, and if that dust, at each wavelength $\lambda$, attenuates\footnote{%
The interstellar dust attenuates the light in the sense that it scatters some of the light out of the line of sight and absorbs some of the light and re-radiates it at longer wavelengths, distributed over a larger solid angle on the sky than that subtended by the star.
The full accounting for the radiative processes involved is way beyond the scope of this \documentname.}
a fraction of the light specified by some optical depth $\tau(\lambda)$, then the generalization of this relationship is
\begin{align}
    f_\lambda(\lambda) &= \frac{1}{4\pi\,D^2}\,L_\lambda(\lambda)\,\exp\left(-\tau(\lambda)\right)\\
    A(\lambda) &\approx 1.086\,\tau(\lambda)~,
\end{align}
where the function $A(\lambda)$ is very similar to the optical depth but with a factor of 1.086 that converts from natural-log units to magnitude units (because $2.5\logten\e\approx 1.086$).

All this means that if the absolute bolometric magnitude of the target star is based on an integral of its spectral luminosity density $L_\lambda(\lambda)$, and the spectral flux density $f_\lambda(\lambda)$ at the top of the atmosphere depends on both the distance to the star and the interstellar attenuation, the relationship between apparent magnitude $m_b$ and absolute magnitude $M_b$ or absolute bolometric magnitude $M_\bol$ must be further complexified to
\begin{align}
    m_b &= M_b + DM + A(b)\label{eq:mA}\\
        &= M_\bol - BC_b + DM + A(b)\\
    A(b) &= -2.5\,\logten\frac{\chi_{bA}}{\chi_{b}}\label{eq:A}\\
    \chi_{bA} &= \int \frac{L_\lambda(\lambda)}{h\,c/\lambda}\,\exp(-\tau(\lambda))\,R_b(\lambda)\,\dd\lambda\\
    \chi_b &= \int \frac{L_\lambda(\lambda)}{h\,c/\lambda}\,R_b(\lambda)\,\dd\lambda ~,
    \end{align}
where $A(b)\geq 0$ is the \emph{extinction}, or a magnitude-units quantity expressing the amount of interstellar dust attenuation,
and $\chi_b,\chi_{bA}$ are hypothetical photon-counting signals you would get if you could photon-count the source with and without dust attenuation.
Recall (from \secref{sec:mag}) the assumption that your device is a photon-counting device.
I apologize that the notation for the infinitesimal-wavelength extinction $A(\lambda)$ is so similar to the finite-bandpass extinction $A(b)$, but it's not my fault!

These relationships---\eqref{eq:mA} and \eqref{eq:A}---constitute a nice definition for the extinction $A(R)$, but it is worth noting that \emph{you need to know a lot of things} in order to compute it accurately:
You need to know the intrinsic spectral luminosity density $L_\lambda(\lambda)$ of the target star, you need to know the amplitude and detailed shape of the attenuation law $\tau(\lambda)$ or $A(\lambda)$, and you need to know the bandpass transmission function $R_b(\lambda)$.
That's a lot, and all of these things are, in general, subject to substantial theoretical and observational uncertainties.
That said, things get simpler as the bandpass $R_b(\lambda)$ gets narrower in wavelength space:
As $R_b(\lambda)$ becomes more concentrated around some central wavelength $\lambda_0$, $A(b)$ approaches $A(\lambda_0)$.
And indeed, much extinction work is done in this limit (and takes on corresponding inaccuracies for very red and very blue stars).
Another way to say all this is that, as the bandpasses in play get wider, more and more information gets lost about the spectral shape of the star and the spectral response of the interstellar medium.

In general the optical-depth and extinction functions $\tau(\lambda)$ and $A(\lambda)$ are larger at shorter wavelengths (although there are exceptions near certain features).
Interstellar dust preferentially absorbs and scatters light at shorter wavelengths.
This means that the introduction of interstellar dust tends to make sources redder, or increases the numerical value of any color indices.
If this is specified to the $B$ and $V$ photometric bands---as it often is---the color index $B-V$ can be expressed in terms of the relationship $\eqref{eq:mA}$
\begin{align}
    B - V &= BC_V - BC_B + A(B) - A(V)\\
          &= (B - V)_0 + E(B - V)\\
    E(B - V) &\equiv A(B) - A(V) ~,
\end{align}
where I have defined the dust-corrected or de-reddened color index $(B-V)_0$,
and the \emph{reddening} or \emph{color excess} $E(B-V)$ in the $B-V$ color index.
Although we could have used any bandpasses we like, and any color indices we like,
a lot of extinction work involves the $B$ and $V$ photometric bandpasses, or translations thereto, so it behooves us to make use of it.

The consequences of these definitions are the following:
The extinctions $A(\lambda)$ and $A(R)$ are zero if there is no interstellar attenuation, and positive otherwise.
They increase with the amount of intervening interstellar dust.
The reddening $E(B-V)$ has the same properties---zero at zero attenuation, and increasing with increasing attenuation---and it usually will for any color index you like, provided that the color index is defined in the usual way, as the shorter-wavelength bandpass magnitude minus the longer-wavelength bandpass magnitude.
This is because dust generally reddens a source!
The sign might change in bandpasses that are near or span a strong feature in the dust attenuation law, such as the 2175-angstrom bump; see, for example \cite{extinction}.

Different kinds of dust---with, say, different particle size distributions---have different kinds of extinction and reddening behaviors as a function of wavelength.
To parameterize this, it is traditional to define a dust reddening law parameter $R(V)$  as
\begin{align}
    R(V) &= \frac{A(V)}{A(B) - A(V)}\label{eq:RV} ~,
\end{align}
which typically has values in the 2 to 3 range \cite{extinction, extinction2}, and varies with environment and redshift.\footnote{%
The reddening-law parameter $R(V)$ should not be confused with the bandpass function $R_b(\lambda)$. I apologize for the over-loaded notation, but this \documentname{} collides multiple literatures.}
This ratio $R(V)$ was originally conceived as the slope of a ``reddening vector'' on a color--absolute-magnitude diagram; the reddening vector points in the direction in the space of $B-V$ color and absolute magnitude $M_V$ that a star would move if the amount of interstellar dust along its line of sight were increased.
This definition \eqref{eq:RV} refers everything to the $B$ and $V$ bands, which is traditional, but not required (and it could be re-done in any bands).
Indeed, the $B$ and $V$ bands are wide enough that this number $R(V)$ cannot be precisely defined without making a choice for a reference spectrum $L_\lambda(\lambda)$ of the target star. This shows that this way of thinking about the dust reddening law is being understood in a narrow-bandpass limit. That's technically approximate, and limits its precision.

One final comment on this theme is that when the bandpasses are narrow, $A()$ and $E()$ increase linearly with the column density of dust along the line of sight.
But when the bandpasses are wide, this is not true, fundamentally because the multiplication (absorption and scattering are multiplicative effects) and integration (the magnitude involves an integral) operators don't commute! This is discussed in detail in \cite{green}, along with a nice result that the bandpass-dependent nonlinearities are proportional to the square of the width of the bandpass.

All this is great, but how can an observer possibly know $A(b)$ or any of these related quantities, such as $R(V)$ or $E(B-V)$?
The answer is different in different contexts.
In some, the investigator has multi-band photometry of reddened and unreddened stars, and it is possible to work out which way dust has moved the reddened ones (because, say, the red-giant branch is very narrow in the relevant direction; see, for a recent example, \cite{dalcanton}).
In others, the investigator has stellar spectra, and reasonable physical models of those spectra that predict the stellar broad-band photometry, to which the observations can be compared (see, for a recent example, \cite{schlafly}).
But it is hard to know precisely the individual reddening or extinction towards any individual star without substantial information or models.

\section{K correction}\label{sec:kcorrect}

I mentioned in \secref{sec:absmag} the counterfactual absurdity in cosmology that often enormous and cosmologically distant objects are described in terms of absolute magnitudes.
If that isn't bad enough:
Cosmologically distant objects are usually redshifted with respect to us, such that the wavelengths of the photons that fall into the transmission function $R_b(\lambda)$ here in the Solar System were shorter when they were emitted in the rest frame of the target.
This leads to another kind of correction, coming from this relativistic Doppler shift.

With colleagues I have discussed this in great detail elsewhere \cite{kcorrect}, but briefly, it is traditional to define a correction $K(z)$ such that
\begin{align}
    m_b &= M_b + DM(z) + A(R) + K(z)\label{eq:K}\\
    K &= -2.5\logten\frac{\xi_{bz}}{\xi_{b0}}\\
    \xi_{bz} &= \int\frac{L_\lambda(\lambda/[1+z])}{([1+z]\,h\,c/\lambda)}\,R_b(\lambda)\,\dd\lambda\\
    \xi_{b0} &= \int\frac{L_\lambda(\lambda)}{(h\,c/\lambda)}\,R_b(\lambda)\,\dd\lambda ~,
\end{align}
where $z$ is the redshift,
it is important that the distance $D$ involved in the distance modulus $DM(z)$ is the luminosity distance \cite{distances},
$K$ has a magnitude-like definition based on hypothetical signals $\xi_{bz}, \xi_{b0}$,
and the hypothetical signal $\xi_{bz}$ involves the spectrum shifted by the Doppler factor $1+z$.

The consequences of these definitions are the following:
The K~correction $K(z)$ vanishes if either the redshift $z$ is zero or else the target source has a spectrum like $\lambda\,L_\lambda(\lambda)=\nu\,L_\nu(\nu)=\text{constant}$ at the relevant wavelengths.
The K-correction is positive and grows with redshift if the target is redder than this in or around the bandpass.
It is negative and grows more negative with redshift if the target is bluer than this.

In some very clever projects, cosmologists use many bandpasses that cover a wide range of wavelength, and then the Doppler shift brings light at some rest-frame wavelength $\lambda$ from one observer-frame bandpass to another as redshift increases.
Thus---unlike how the definition \eqref{eq:K} is written---the bandpass in which the apparent magnitude is measured is often different from the bandpass in which the absolute magnitude is inferred. Then the $K(z)$ term gets more complex.
That is out of scope here, but it is precisely the scope of our previous writing on this subject \cite{kcorrect}.
I recommend that as a jumping-off point if you are working on such projects.

One amusing point about the K correction and redshift is that, when similar galaxies or quasars are observed at higher and higher redshifts, the higher-redshift objects are not in general \emph{redder} than the lower-redshift objects.
Whether objects in some class generally get redder or bluer as redshift increases depends on the detailed spectral shape of the objects in the class, the redshift ranges in question, and the particular bandpasses being used.

\section{Real-world measurement}\label{sec:practice}

Everything above has been about the \emph{theoretical expectations} for magnitudes and corrections, which you can compute under the strong assumptions that you know your bandpass and the spectrum of your target star.
But the whole original point of the magnitude system is that you can \emph{measure} apparent magnitudes without knowing any of those things.
How do these measurements work in practice?

The first comment to make is that it is rare now to actually compare your target star \emph{directly} with the fundamental standard star.
That is, it is rare to point the same instrument and telescope at the target star and at the zeroth-magnitude Vega.
This is because, in general, contemporary science targets are faint, and detectors don't have the dynamic range to observe the target star and Vega in the same mode.
So in practice the usual approach is to calibrate the science target to a set of fainter secondary standards, which have been calibrated relative to the standard (see, for example, \cite{landolt}).
That's great.
We illustrated explicitly how calibrated secondary standards are used to estimate magnitudes in \eqref{eq:secondary} in \secref{sec:mag}.

When you are lucky and the sky is dense with secondary stars, you might even have secondaries in your field of view as you observe your target star (this is often true when the secondaries come from a large, deep sky survey, such as \textsl{SDSS} or \textsl{2MASS}, as we note below).
If you don't have secondaries in your field of view, then (if you want good magnitudes) you have to make multiple measurements over the night, keeping track of the airmass (see \noteref{note:airmass} for the definition of airmass) at which the target-star and secondary-star measurements are made.
With a regression on the detector signal as a function of airmass it is possible to correct both the target-star and secondary-star signals to unit airmass before computing the magnitude.

Of course if you are doing very precise work, it will turn out that the bandpass at which you are doing your observations don't match exactly the bandpasses that were used to measure the secondary standard stars.
You have to match these bandpasses, and you usually do this by a regression or interpolation, in which you learn functional forms for \emph{synthetic magnitudes}.
That is, you find a linear (or nonlinear) combination of the magnitudes and color indices that have been measured for the secondary calibrators to create an approximation to the magnitudes for the secondary stars that \emph{would have been observed} if the secondary stars had been compared to the primary standard through the bandpass in which you are working.
This is complicated! But it is also data-driven: If you have enough measured secondary standards and they span enough range in temperature, the construction of these synthetic magnitudes is generally possible.

If you want to produce not just apparent magnitudes for your target stars, but also some kind of uncertainty estimate, there are several things to consider.
The first is that the magnitude is a logarithm of a ratio of counts.
If each of these counts is subject to a quasi-Gaussian noise (perhaps a sum of high-count Poisson noise, Gaussian sky noise, and some kind of detector read noise), the logarithm of the ratio of them will not, in general, have quasi-Gaussian noise.
In particular, the noise on a magnitude will tend to be ``fat-tailed'', and also skew.
The next thing to consider is that there are many systematic-noise or bias contributions to the measurment, including from mis-estimation of differential atmospheric extinction, inconsistency in the synthetic-magnitude system in use, and detector nonlinearities.
A full discussion of the noise in a measurement like this is out of scope.

Nowadays, the secondary standards are often not really standards at all:
Often the calibration of new photometry or new imaging is made with respect to prior imaging of that same patch of sky, from a large survey (such as \textsl{2MASS} \cite{2mass} or \textsl{SDSS} \cite{sdss} or \textsl{PanSTARRS} \cite{panstarrs}).
Once again, synthetic magnitudes must be created, but if the prior imaging overlaps your target-star field, you can often make the comparisons required for your apparent magnitude measurements within your science imaging itself, without taking additional calibration frames.
That has really changed astrophysics enormously.\footnote{This point was made in multiple contributions, now lost in time, at the meeting \textit{Landolt Standards and 21st Century Photometry} organized at Louisiana State University in 2015 by Pagnotta and Clayton in honor of Arlo Landolt.}
And it often makes it possible that you have good calibrators in every image you take, or a large fraction of them.

If you want to create your own secondary standards, because, say, you need standards in a new magnitude range, or, say, you need standards that are in some particular temperature or metallicity ranges, there are a host of considerations.
Any complete discussion of those considerations is out of scope here, but the issues include that noise sources, sky levels, and even detector properties \cite{brighterfatter} can vary with the incident intensity of the star.
Calibrating secondary standards many magnitudes away from the primary standard requires extremely detail-oriented work.

The very most precisely calibrated systems in astrophysics now are \emph{self-calibrated}.
That means that their fundamental calibration comes not (primarily) from comparison to external, calibrated secondary standards, but their fundamental calibration comes (primarily) from internal self-consistency.
This was how we calibrated the DR8 imaging of the \textsl{SDSS} \cite{sdss}, and it was far more precise than the best calibrations we could achieve by comparisons to external calibrators \cite{ubercal}.
Self-calibration is the method of choice for most cosmology missions now, and it will be the method of choice for all future large optical photometric projects as well, I expect.
Self-calibration can't tell you everything about your calibration, but it can often tell you everything except one overall scale or offset.

\section{Comments and discussion}\label{sec:discussion}

The writing of this \documentname{} was originally inspired by a few conversations I have had about computing expectations for apparent magnitudes and bolometric corrections with experienced scientists.
In these conversations, sometimes the inclusion of the $(h\,c/\lambda)$ factor in the integral \eqref{eq:counts} led to expressions of surprise or even disbelief.
It is easy to naively assume that a magnitude is an integral of an energy, especially since the filter curves are measured using intensity measurements.
It is not!
If there is one important take-away from this \documentname, it is that it is important to perform the integrals carefully and with the correct photon-counting ``kernel'' as it were.
And, in this context, it is important to see the transmission curve $R_b(\lambda)$ as the expected contribution to the detector signal from a \emph{photon} at the top of the atmosphere.

For an example of how upsetting these integral kernels are, consider the expressions for the bolometric corrections in \secref{sec:bc}.
These expressions mix up logarithms of ratios of integrals that do and don't have the energy-per-photon factor.
That's not good, intellectually.
But it is required, unfortunately, because most kinds of optical detectors are photon-counting devices, and yet most kinds of theories predict integrated energy outputs of stars, not integrated \emph{photon} outputs.

On a related side note:
If you actually do these expectation integrals (on a computer, say), you will find that results can depend on the integration scheme and stepsize.
Stellar spectra are highly featured, and the features can be narrow and sharp.
Integrate with care.

As noted above in \secref{sec:transmission}, it is not usually a good idea to think of a wide photometric bandpass as having an ``effective wavelength'' or center, as any such measure depends strongly on the spectrum of the target star (and also the standard star).
Related to this, it is generally true that the analysis or prediction or interpretation of apparent magnitudes (and related quantities like color indices) gets far simpler when \emph{bandpasses get narrow}, and gets far more complicated when \emph{bandpasses get broader}.
The concept of the effective wavelength of a bandpass is more dependent on the detailed stellar spectrum the broader the bandpass, and is more uniform across stellar spectra the narrower the bandpass.
For example, as we note in \secref{sec:extinction}, the effect of dust attenuation is more dependent on the detailed stellar spectrum as the bandpass gets broader.
Fundamentally this is because the effect of dust involves a multiplication (by the dust law) followed by an integral (over the bandpass), and multiplication and integration don't commute.
One might ask: Then why not just work in narrow bandpasses?
Indeed, some projects have taken this route \cite{combo17}.
But in the end, the choice of a bandpass is a balance between signal-to-noise (which pushes projects to broader bandpasses) and spectral specificity or resolution (which pushes projects to narrower bandpasses).
Since astronomy is focused, most of the time, on photon-limited science, we often choose broader bandpasses than we would like, at least from an astrophysics-theory perspective.

Bandpasses involve the atmosphere, which varies in time in dust content, humidity, pressure, and temperature.
Bandpasses also involve the detector sensitivity or efficiency, which vary with pixel, and even intra-pixel \cite{spitzerpixels}.
Thus every photometric measurement has a bandpass that depends on time (weather) and position in the telescope focal plane (and lots of other things, such as mirror re-aluminization \cite{ubercal}).
In some real sense, every photometric measurement ever made has been made through a unique, \emph{unrepeatable bandpass}!
No two nights, no two pixels, no two observations, no two instruments will ever have exactly the same bandpasses.
And if they don't have the same bandpasses, they don't have the same magnitude systems at the same level.
That's a challenge. But that's reality, and it limits our precision in the end.

In everything above, I considered only the magnitudes of point sources (or compact sources anyway).
Isn't this odd, since really it is the \emph{intensity} that is the fundamental quantity in electromagnetism?\footnote{The intensity is energy per wavelength per time per area per solid angle. It is related to the phase-space density of photons, which is what is preserved along geodesic trajectories in a transparent Universe. Fluxes are not what's preserved. See, for example, gravitational-lensing magnifications, which preserve intensity but not fluxes \cite{lensing}.}
Of course astronomers do measure intensities very often---optical astronomers often call these ``surface brightnesses''---and they often use the same techniques described above, but with modifications to account for the distribution of the light in the detector image plane.
Optical astronomers often use units of ``magnitudes per square arcsecond'' for such measurements, which is an abomination and ought to be avoided.
After all, magnitudes don't increase linearly with solid angle!
In general there are issues that arise when one ties an intensity measurement (from a bright patch in your imaging, say) to a flux measurement (from an unresolved secondary standard star in your imaging), because these things have very different units and observational footprints.
One indication of danger is that there are conceptual puzzles and challenges with such measurements involving the point-spread function, pixel areas, and flat-field maps (the reader is referred to, for example, \cite{bernstein} for examples of the challenges).
That said, I don't have a worked-out alternative recommendation at present.

In the long run, if we want to make extremely repeatable photometric measurements from the ground, and if we want those measurements to be tied to an accurately known and stable standard source, it might make sense to \emph{fly} a standard star.
That is, it might be a good idea to put a stable black-body (or set of black-bodies) onto a spacecraft and place it onto some useful orbit.
This would be the next generation for experiments like the one in which various calibrated blackbodies were placed on a tower near the Palomar 5-m telescope \cite{okeschild}.
Space-based calibration sources have been proposed \cite{wacky}, and indeed the extremely precise \textsl{FIRAS} measurements \cite{firas} of the cosmic microwave background intensity---which are perhaps the most accurate\footnote{%
In this sentence and the next I am intentionally using the word ``accurate''; I make a strong distinction between precision and accuracy.}
astronomical measurements of intensity ever made---were made by making on-board comparisons to an on-board standard intensity source \cite{firascalibrator}.
If astronomical photometry has a high-accuracy future---and not just a high-precision future---something of this nature might be required.

\begin{wider}
\bibliographystyle{plain}
\raggedright
\bibliography{mags}

\begin{thebibliography}{10}

\bibitem{sdss}
H.~{Aihara} et~al.
\newblock The {Eighth Data Release} of the {Sloan Digital Sky Survey}: First
  data from {SDSS-III}.
\newblock {\em The Astrophysical Journal Supplement Series}, 193(2):29, 2011.

\bibitem{wacky}
J.~{Albert}.
\newblock Satellite-mounted light sources as photometric calibration standards
  for ground-based telescopes.
\newblock {\em The Astronomical Journal}, 143(1):8, 2012.

\bibitem{brighterfatter}
Pierre Antilogus, P~Astier, P~Doherty, A~Guyonnet, and N~Regnault.
\newblock The brighter-fatter effect and pixel correlations in {CCD} sensors.
\newblock {\em Journal of Instrumentation}, 9(03):C03048, 2014.

\bibitem{kepler}
N.~M. {Batalha} et~al.
\newblock Planetary candidates observed by {Kepler}. {III}. analysis of the
  first 16 months of data.
\newblock {\em The Astrophysical Journal Supplement Series}, 204(2):24, 2013.

\bibitem{bernstein}
R.~A. {Bernstein}, W.~L. {Freedman}, and B.~F. {Madore}.
\newblock The first detections of the extragalactic background light at 3000,
  5500, and 8000 angstroms. {II}. measurement of foreground {Zodiacal} light.
\newblock {\em The Astrophysical Journal}, 571(1):85--106, 2002.

\bibitem{dalcanton}
J.~J. {Dalcanton} et~al.
\newblock The {Panchromatic Hubble Andromeda Treasury}. {VIII}. {A} wide-area,
  high-resolution map of dust extinction in {M31}.
\newblock {\em The Astrophysical Journal}, 814(1):3, 2015.

\bibitem{doi}
M.~{Doi} et~al.
\newblock Photometric response functions of the {Sloan Digital Sky Survey}
  imager.
\newblock {\em The Astronomical Journal}, 139(4):1628, 2010.

\bibitem{extinction}
E.~L. {Fitzpatrick}, D.~{Massa}, K.~D. {Gordon}, R.~{Bohlin}, and G.~C.
  {Clayton}.
\newblock An analysis of the shapes of interstellar extinction curves. {VII}.
  {Milky Way} spectrophotometric optical-through-ultraviolet extinction and its
  {R}-dependence.
\newblock {\em The Astrophysical Journal}, 886(2):108, 2019.

\bibitem{gaia}
{Gaia Collaboration}, T.~{Prusti}, et~al.
\newblock The {Gaia} mission.
\newblock {\em Astronomy \& Astrophysics}, 595:A1, 2016.

\bibitem{acs}
S.~Gonzaga et~al.
\newblock {\em {ACS} Instrument Handbook, Version 6.0}.
\newblock Space Telescope Science Institute, 2005.

\bibitem{green}
G.~M. {Green} et~al.
\newblock Data-driven stellar models.
\newblock {\em The Astrophysical Journal}, 907(1):57, 2021.

\bibitem{hearnshaw}
J.~B. {Hearnshaw}.
\newblock {\em The Measurement of Starlight: Two Centuries of Astronomical
  Photometry}.
\newblock Cambridge, 1996.

\bibitem{distances}
D.~W. {Hogg}.
\newblock Distance measures in cosmology.
\newblock {\em arXiv}, astro-ph/9905116, 1999.

\bibitem{kcorrect}
D.~W. {Hogg}, I.~K. {Baldry}, M.~R. {Blanton}, and D.~J. {Eisenstein}.
\newblock The {K} correction.
\newblock {\em arXiv}, astro-ph/0210394, 2002.

\bibitem{vegadust}
W.~S. {Holland} et~al.
\newblock Submillimetre images of dusty debris around nearby stars.
\newblock {\em Nature}, 392(6678):788--791, 1998.

\bibitem{spitzerpixels}
J.~G. {Ingalls}, J.~E. {Krick}, S.~J. {Carey}, S.~{Laine}, J.~A. {Surace},
  W.~J. {Glaccum}, C.~C. {Grillmair}, and P.~J. {Lowrance}.
\newblock Intra-pixel gain variations and high-precision photometry with the
  {Infrared Array Camera (IRAC)}.
\newblock In M.~C. {Clampin}, G.~G. {Fazio}, H.~A. {MacEwen}, and Jr.
  {Oschmann}, J.~M., editors, {\em Space Telescopes and Instrumentation 2012:
  Optical, Infrared, and Millimeter Wave}, volume 8442 of {\em Society of
  Photo-Optical Instrumentation Engineers (SPIE) Conference Series}, page
  84421Y, 2012.

\bibitem{lsstfilters}
\v{Z}. Ivezi\'c et~al.
\newblock {LSST}: From science drivers to reference design and anticipated data
  products.
\newblock {\em The Astrophysical Journal}, 873(2):111, 2019.

\bibitem{jordi}
C.~{Jordi} et~al.
\newblock {Gaia} broad band photometry.
\newblock {\em Astronomy \& Astrophysics}, 523:A48, 2010.

\bibitem{panstarrs}
N.~{Kaiser} et~al.
\newblock {The Pan-STARRS} wide-field optical/{NIR} imaging survey.
\newblock In L.~M. {Stepp}, R.~{Gilmozzi}, and H.~J. {Hall}, editors, {\em
  Ground-based and Airborne Telescopes III}, volume 7733 of {\em Society of
  Photo-Optical Instrumentation Engineers (SPIE) Conference Series}, page
  77330E, 2010.

\bibitem{landolt}
A.~U. {Landolt}.
\newblock {UBVRI} photometric standard stars in the magnitude range $11.5 < {V}
  < 16.0$ around the {Celestial Equator}.
\newblock {\em The Astronomical Journal}, 104:340, 1992.

\bibitem{2015b2}
E.~E. {Mamajek} et~al.
\newblock {IAU} 2015 resolution {B2} on recommended zero points for the
  absolute and apparent bolometric magnitude scales.
\newblock {\em arXiv}, 1510.06262, 2015.

\bibitem{firas}
J.~C. {Mather} et~al.
\newblock Measurement of the cosmic microwave background spectrum by the {COBE
  FIRAS} instrument.
\newblock {\em The Astrophysical Journal}, 420:439, 1994.

\bibitem{firascalibrator}
J.~C. {Mather}, D.~J. {Fixsen}, R.~A. {Shafer}, C.~{Mosier}, and D.~T.
  {Wilkinson}.
\newblock Calibrator design for the {COBE} {Far-Infrared Absolute
  Spectrophotometer (FIRAS)}.
\newblock {\em The Astrophysical Journal}, 512(2):511, 1999.

\bibitem{combo17}
K.~{Meisenheimer} and C.~{Wolf}.
\newblock {COMBO-17}: Fifty thousand galaxies at a glance.
\newblock {\em Astronomy and Geophysics}, 43(3):3.15--3.21, 2002.

\bibitem{ab}
J.~B. {Oke} and J.~E. {Gunn}.
\newblock Secondary standard stars for absolute spectrophotometry.
\newblock {\em The Astrophysical Journal}, 266:713, 1983.

\bibitem{okeschild}
J.~B. {Oke} and R.~E. {Schild}.
\newblock The absolute spectral energy distribution of {Alpha Lyrae}.
\newblock {\em The Astrophysical Journal}, 161:1015, 1970.

\bibitem{ubercal}
N.~{Padmanabhan} et~al.
\newblock An improved photometric calibration of the {Sloan Digital Sky Survey}
  imaging data.
\newblock {\em The Astrophysical Journal}, 674(2):1217--1233, 2008.

\bibitem{extinction2}
J.~E.~G. {Peek} and D.~{Schiminovich}.
\newblock Ultraviolet extinction at high {Galactic} latitudes.
\newblock {\em The Astrophysical Journal}, 771(1):68, 2013.

\bibitem{hstcmd}
Ata {Sarajedini} et~al.
\newblock The {ACS Survey} of {Galactic} globular clusters. {I}. {Overview} and
  clusters without previous {Hubble Space Telescope} photometry.
\newblock {\em The Astronomical Journal}, 133(4):1658--1672, 2007.

\bibitem{schlafly}
E.~F. {Schlafly}, J.~E.~G. {Peek}, D.~P. {Finkbeiner}, and G.~M. {Green}.
\newblock Mapping the extinction curve in {3D}: {Structure} on kiloparsec
  scales.
\newblock {\em The Astrophysical Journal}, 838(1):36, 2017.

\bibitem{lensing}
P.~{Schneider}, J.~{Ehlers}, and E.~E. {Falco}.
\newblock {\em Gravitational Lenses}.
\newblock Springer, 1992.

\bibitem{2mass}
M.~F. {Skrutskie} et~al.
\newblock {The Two Micron All Sky Survey (2MASS)}.
\newblock {\em The Astronomical Journal}, 131(2):1163--1183, 2006.

\bibitem{spitzer}
M.~W. {Werner} et~al.
\newblock {The Spitzer Space Telescope} mission.
\newblock {\em The Astrophysical Journal Supplement Series}, 154(1):1--9, 2004.

\end{thebibliography}
\end{wider}

\end{document}